\documentstyle[aps,multicol,prb,eqsecnum]{revtex}
\renewcommand{\narrowtext}{\begin{multicols}{2} \global\columnwidth20.5pc}
\renewcommand{\widetext}{\end{multicols} \global\columnwidth42.5pc}
\input{epsf.tex}
\def\inseps#1#2{\def\epsfsize##1##2{#2##1} \centerline{\epsfbox{#1}}}
\def\top#1{\vskip #1\begin{picture}(290,80)(80,500)\thinlines \put(
65,500){\line( 1, 0){255}}\put(320,500){\line( 0, 1){
5}}\end{picture}}
\def\bottom#1{\vskip #1\begin{picture}(290,80)(80,500)\thinlines \put(
330,500){\line( 1, 0){255}}\put(330,500){\line( 0, -1){
5}}\end{picture}}

\begin{document}
\draft
\title{
Evolution from BCS Superconductivity to Bose Condensation:
Analytic Results for the crossover in three dimensions}

\author{M. Marini$^{(1)}$, F. Pistolesi,$^{(1,2)}$\cite{FP} 
and G.C. Strinati$^{(1)}$}

\address{$(1)$ Dipartimento di Matematica e Fisica, Sezione INFM,
Universit\`a di Camerino, I-62032 Camerino, Italy}

\address{$(2)$ Scuola Normale Superiore, Sezione INFM, I-56126 Pisa, Italy}

\date{11 March 1997}

\maketitle
\begin{abstract}
We provide an analytic solution for the mean-field equations 
and for the relevant physical quantities at the Gaussian level,
in terms of the 
complete elliptic integrals of the first and second kinds,
for the crossover problem from BCS 
superconductivity to Bose-Einstein condensation
of a {\it three}-dimensional system of free fermions interacting 
via an attractive contact potential at zero temperature.
This analytic solution enables us to follow the evolution
between the two limits in a particularly simple and 
transparent way, as well as to verify the absence of singularities during 
the evolution.
\end{abstract}


\narrowtext

\section{Introduction}

The interest in the crossover from BCS superconductivity
to Bose-Einstein (BE) condensation has quite recently 
increased both experimentally and theoretically after the appearance
of new angular resolved photoemission data on underdoped cuprate 
samples, which indicate the persistence 
of a gap in the single-particle excitation spectrum at temperatures
well above the superconducting critical temperature.\cite{R1,R2,R3}
Previous to these data the suggestion that superconductivity
in cuprates (as well as in other ``exotic'' materials) might require 
an intermediate approach between the two (BCS and BE) limits 
has been emphasized especially by Uemura.\cite{R4}

From the theoretical point of view, evolution 
from BCS superconductivity to BE condensation has been 
considered originally by Nozi\`eres and Schmitt-Rink,\cite{R5}
following the pioneering works by 
Eagles\cite{R6} and Leggett.\cite{R7}
More recently, the crossover problem was addressed in Refs.
\onlinecite{R8,R9,R10}, 
prompted by the experimental suggestion that the superconducting 
coherence length in cuprates is considerably
shorter than in conventional superconductors. 
In all these works known results were recovered analytically in the two (BCS
and BE) limits. However, no analytic result was obtained  for the most 
interesting crossover region, which had thus to be treated numerically
(with the sole exception of the two-dimensional case considered 
in Ref. \onlinecite{R11}).

In this paper we provide an analytic solution 
in the {\em three dimensional\/} case covering the 
{\em whole\/} crossover region for a number of physical 
quantities evaluated at the mean-field level and with the 
inclusion of Gaussian fluctuations, by considering a system 
of fermions in free space at zero temperature
mutually interacting via an attractive {\em contact\/} 
potential. For this case we are, in fact, able to express these 
physical quantities in terms of the complete elliptic integrals of 
the first and second kinds, whose analytic properties and numerical 
values are extensively tabulated.\cite{R12} 
Our solution enables us to interpolate in a rigorous fashion 
between the two (BCS and BE) limits, thus avoiding the problems 
which occur with a full numerical approach. 
Although our analytical solution has been unavoidably obtained 
with simplifying assumptions
(namely, at zero temperature and using a free-particle
dispersion relation and a contact interaction potential), 
it might nevertheless be regarded as a reference solution with 
which numerical solutions obtained for more complicated 
cases could be compared.

The plan of the paper is as follows. In Section \ref{II} 
we express a number of mean-field quantities (such as the chemical
potential $\mu$, the gap parameter $\Delta$, the pair-correlation
length $\xi_{pair}$, and the bound-state energy $\epsilon_o$)
in terms of the complete elliptic integrals of the first and 
second kinds with argument 
$\kappa^2 = (\sqrt{1+x_o^2}+x_o)/(2\sqrt{1+x_o^2})$
[where $x_o=\mu/\Delta$] ranging from 0 (BE limit) to 1 (BCS limit).
In Section \ref{III} our analysis is extended to quantities 
such as the phase coherence length $\xi_{phase}$ and 
the sound velocity $s$, by considering the Gaussian fluctuations about
the mean field. To keep the presentation compact, the 
properties of the elliptic integrals used in our treatment are 
summarized in Appendix \ref{A}. In Appendix \ref{B} 
we report for the sake of comparison the solution for the 
two-dimensional case given previously by Ref. \onlinecite{R11}.

\section{Analytic Results at the Mean-fied Level}
\label{II}

We consider the Hamiltonian ($\hbar=1$)
\begin{eqnarray}
H &=& \sum_\sigma \int \! d{\bf r} 
	\, \psi_\sigma^\dag({\bf r})
	\left(-{{\bf \nabla}^2 \over 2m} -\mu\right)
	\psi_\sigma({\bf r})
	\nonumber \\
	&& + g \int\!d{\bf r}\, 
	\psi_\uparrow^\dag({\bf r}) \psi_\downarrow^\dag({\bf r})
	\psi_\downarrow^{\phantom{\dag}}({\bf r}) 
	\psi_\uparrow^{\phantom{\dag}}({\bf r})
	\label{2.1}	
\end{eqnarray}
where $\psi_\sigma({\bf r})$ is the fermionic field operator 
with spin projection $\sigma$, $m$ is the fermionic (effective) mass,
and $g=V \Omega$ is the strength of the short-range (contact)
potential between fermions with $V<0$ ($\Omega$ being the 
volume occupied by the system).

At the zero temperature, the mean-field equations for the 
gap parameter $\Delta$ and the chemical potential $\mu$
are obtained by a suitable decoupling of the Hamiltonian (\ref{2.1})
and are given by\cite{R13}
\begin{equation}
-{1\over V} = \sum_{\bf k} {1\over 2 E_{\bf k}}
\label{2.2}
\end{equation}
\begin{equation}
	n = {N\over \Omega} = {2\over \Omega} \sum_{\bf k} v_{\bf k}^2
\label{2.3}
\end{equation}
where $\bf k$ is the wave vector, $N$ the total number 
of fermions, and 
\begin{equation}
\xi_{\bf k} = {{\bf k}^2 \over 2 m} - \mu,
\ E_{\bf k}=\sqrt{\xi_{\bf k}^2 + \Delta^2},
\ v_{\bf k}^2 = {1\over 2}\left(1-{\xi_{\bf k} \over E_{\bf k}}\right).
\label{2.4}
\end{equation}
Owing to our choice of a contact potential,
Eq. (\ref{2.2}) diverges in the ultraviolet (both in two and three dimensions)
and requires a suitable regularization. In three dimensions 
it is common practice to introduce the scattering amplitude
$a_s$ defined via the equation\cite{R8,R9,R10}
\begin{equation}
	{m \over 4 \pi a_s} = {1\over \Omega V} 
	+ {1\over \Omega} \sum_{\bf k} {m \over {\bf k}^2} \quad,
	\label{2.5}
\end{equation}
where the divergent sum on the right-hand side of Eq.(\ref{2.5})
results in a finite value of $a_s$ by letting $V\rightarrow 0$
in a suitable way. Subtracting  Eq. (\ref{2.5}) from Eq. (\ref{2.2})
we obtain
\begin{equation}
	-{m \over 4 \pi a_s} = {1\over \Omega} \sum_{\bf k} 
	\left({1\over 2 E_{\bf k}} - {m \over {\bf k}^2} \right)
	\label{2.6}
\end{equation}
which has to be solved simultaneously with Eq. (\ref{2.3}) to 
determine $\Delta$ and $\mu$ as functions of $a_s$.

In three dimensions it is convenient to introduce the following 
dimensionless quantities
\begin{equation}
\left\{
\begin{array}{rclrcl}
 x^2 &=&\displaystyle {k^2 \over 2m} {1\over \Delta}\quad, & x_o 
		&=& \displaystyle {\mu\over \Delta} \quad ,\\
&&&&&\\
\xi_x &=& \displaystyle {\xi_k \over \Delta} = x^2-x_o\quad, 
& E_x &=& \displaystyle {E_k \over \Delta} = \sqrt{\xi_x^2+1} 
\end{array}
\right.
\label{2.7}
\end{equation}
with $k=|{\bf k}|$, and express Eqs. (\ref{2.3}) and (\ref{2.6})
in the form
\begin{mathletters}
\begin{eqnarray}
-{1\over a_s} &=& {2\over \pi} (2m \Delta)^{1/2} I_1(x_o)
\label{2.8a}\\
n &=& {1\over 2 \pi^2} (2m \Delta)^{3/2} I_2(x_o)
\label{2.8b}
\end{eqnarray}
\label{2.8}
\end{mathletters}
where 
\begin{eqnarray}
	I_1(x_o) &=& \int_0^\infty\!\!\!\!dx \, x^2
	\left({1\over E_x} - {1\over x^2}\right)
	\label{2.9} \\
	I_2(x_o) &=& \int_0^\infty\!\!\!\! dx \, x^2
	\left(1 -{\xi_x \over E_x}\right)
	\quad . 
	\label{2.10}
\end{eqnarray}
The integrals (\ref{2.9}) and (\ref{2.10}) 
were originally considered by 
Eagles,\cite{R6} who evaluated them numerically as functions 
of the crossover parameter $x_o$ ranging from $-\infty$ (BE limit)
to $+\infty$ (BCS limit). We shall soon show that the 
integrals (\ref{2.9}) and (\ref{2.10}) can actually be evaluated 
in a closed form for all values of $x_o$. Note that 
$I_2(x_o)\geq 0$ while $I_1(x_o)$ can take both signs.

Before proceeding further, it is convenient to render Eqs. (\ref{2.8})
dimensionless by introducing the Fermi energy 
$\epsilon_F = k_F^2/2m = (3\pi^2 n)^{2/3}/2m$ where $k_F$ is 
the Fermi momentum. In this way Eq. (\ref{2.8b}) becomes
\begin{equation}
	{\Delta\over \epsilon_F} 
	= 
	\left[{2\over 3 I_2(x_o)}\right]^{2/3}
	\label{2.11}
\end{equation}
and Eq. (\ref{2.8a}) reduces to 
\begin{equation}
{1\over k_F a_s} = - {2\over \pi} \left[2 \over 3 I_2(x_o)\right]^{1/3}
	I_1(x_o)\quad .
	\label{2.12}
\end{equation}
Note that the right-hand sides of Eqs. (\ref{2.11}) and (\ref{2.12})
depend on $x_o$ only. Equation (\ref{2.12}) can thus be inverted to obtain
$x_o$ as a function of $k_F a_s$; from Eq. (\ref{2.11}) 
and from $\mu/\epsilon_F = x_o \Delta/\epsilon_F$ one can then 
obtain the two parameters $\Delta/\epsilon_F$ and $\mu/\epsilon_F$
as functions of $k_F a_s$. 

Alternatively, one can use $k_F \xi_{pair}$ as the independent variable
in the place of $k_F a_s$, where $\xi_{pair}$ is the 
characteristic length for pair correlation given by 
the following expression in three dimensions at the mean-field level:\cite{R10}
\begin{equation}
\xi_{pair}^2 = {1\over m^2} 
	{\displaystyle
	 \int_0^\infty\!\!\!\!dk (k^4\xi_k^2/E_k^6) \over 
	\displaystyle \int_0^\infty \!\!\!\! dk (k^2/E_k^2)
	}
	=
	{2\over m \Delta}
	{I_3(x_o) \over I_4(x_o)}
\label{2.13}
\end{equation}
where
\begin{eqnarray}
	I_3(x_o) &=& \int_0^\infty \!\!\!\! dx {x^4 \xi_x^2 \over E_x^6}
	\label{2.14}\\
	I_4(x_o) &=& \int_0^\infty \!\!\!\! dx {x^2\over E_x^2}
	\label{2.15}
\end{eqnarray}
are two additional integrals expressed in terms 
of the quantities (\ref{2.7}). Contrary to the integrals $I_1$ and 
$I_2$, the integrals $I_3$ and $I_4$  are elementary and can be 
evaluated via the residues technique. One obtains:
\begin{eqnarray}
	I_3(x_o) &=& {\pi \over 16} {x_1(1+x_1^4) \over (1+x_o^2)^{1/2}}
	\label{2.16}\\
	I_4(x_o) &=& {\pi \over 2} x_1
	\label{2.17}
\end{eqnarray}
with the notation
\begin{equation}
	x_1^2 = {\sqrt{1+x_o^2}+x_o \over 2}
	\quad . \label{2.18}
\end{equation}
Making use of Eq. (\ref{2.11}) we obtain eventually
\begin{equation}
	(k_F \xi_{pair})^2 
	= {\epsilon_F \over 2 \Delta} 
	{(1+x_1^4) \over (1+x_o^2)^{1/2} }
	= {(1+x_1^4) \over 2 (1+x_o^2)^{1/2}}
	\left[ {3 I_2(x_o) \over 2}\right]^{2/3}.
\label{2.19}
\end{equation}
In this way Eq. (\ref{2.12}) can be dropped in favor of Eq. (\ref{2.19}),
which can be inverted to obtain $x_o$ as a function of $k_F \xi_{pair}$.

There remains to evaluate the integrals $I_1(x_o)$ entering 
Eq. (\ref{2.12}) and $I_2(x_o)$ entering Eqs. (\ref{2.11}) and 
(\ref{2.19}). To 
this end, we introduce the auxiliary integrals
\begin{eqnarray}
	I_5(x_o) &=& \int_0^\infty\!\!\!\! dx {x^2 \over E_x^3}\quad,
	\label{2.20} \\
	I_6(x_o) &=& \int_0^\infty\!\!\!\! dx {x^2 \xi_x \over E_x^3}
	\quad ,
	\label{2.21}
\end{eqnarray}
such that 
\begin{eqnarray}
	I_1(x_o) &=& 2\left(x_o I_6(x_o)-I_5(x_o)\right)
	\label{2.22}\\
	I_2(x_o) &=& {2\over 3} \left(x_o I_5(x_o) + I_6(x_o)\right)
	\label{2.23}
\end{eqnarray}
after integration by parts. The auxiliary integrals
$I_5(x_o)$ and $I_6(x_o)$ can, in turn, be expressed as linear 
combinations of the complete elliptic integrals of the first 
[$F({\pi\over 2},\kappa)$] and second [$E({\pi\over 2},\kappa)$] kinds.
Reduction of a generic integral of the elliptic kind to the normal
Legendre's form has been treated at length in the literature.\cite{R14}
Here we proceed as follows. Integration by parts gives for 
$I_6(x_o)$ [cf. Eq. (\ref{2.21})]:
\begin{eqnarray}
	I_6(x_o) 
	&=& 
	-{1\over2} \int_0^\infty\!\!\!\! dx\, x 
	{d \hphantom{x}\over dx}
	{1\over E_x} = {1\over 2} \int_0^\infty\!\!\!\!dx {1\over E_x}
	\nonumber \\
	&=& 
	{1\over 2} \int_0^\infty\!\!\!\! \,dx 
	{1\over (x^4 -2x_o x^2 + x_o^2 +1)^{1/2}}
	\nonumber \\
	&=& 
	{1\over 2 (1+x_o^2)^{1/4}} 
	F({\pi\over 2}, \kappa)
	\label{2.24}
\end{eqnarray}
where in the last line use has been made of the results of Appendix A 
 and where ($0\leq \kappa^2 < 1$)
\begin{equation}
	\kappa^2 
	= 
	{x_1^2 \over (1+x_o^2)^{1/2}}
	\label{2.25}
\end{equation}
with $x_1$ given by Eq. (\ref{2.18}). For $I_5(x_o)$ we obtain 
instead [cf. Eq. (\ref{2.20})]:
\begin{eqnarray}
	I_5(x_o) 
	&=& 
	\int_0^\infty \!\!\!\! dx 
	{x^2 \over (x^4 - 2 x_o x^2 + x_o^2 +1)^{3/2} }
	\nonumber \\
	&=& 
	(1+x_o^2)^{1/4} 
	E({\pi\over 2},\kappa) - {1\over 4 x_1^2(1+x_o^2)^{1/4}} 
	F({\pi\over 2}, \kappa) \nonumber \\
	\label{2.26}
\end{eqnarray}
where again use has been made of the results of Appendix A.

In conclusion, we obtain for the quantities of interest 
[cf. Eqs. (\ref{2.11}), (\ref{2.12}), and (\ref{2.19})]:
\begin{eqnarray}
	{\Delta \over \epsilon_F} 
	&=& 
	{1\over (x_o I_5(x_o) + I_6(x_o))^{2/3}}
	\label{2.27}
	\\
	{\mu \over \epsilon_F} 
	&=& 
	{\mu \over \Delta} {\Delta \over \epsilon_F} 
	= {x_o \over (x_o I_5(x_o) + I_6(x_o))^{2/3}}
	\label{2.28}
	\\
	{1\over k_F a_s} 
	&=& 
	- {4\over \pi} {(x_o I_6(x_o) - I_5(x_o)) 	
	\over (x_o I_5(x_o) + I_6(x_o))^{1/3}}
	\label{2.29}
	\\
	k_F \xi_{pair} 
	&=& 
	\left({1 + x_1^4 \over 2}\right)^{1/2} 
	{ (x_o I_5(x_o)+I_6(x_o))^{1/3} \over (1+x_o^2)^{1/4}}
	\label{2.30}	
\end{eqnarray}
with $I_5(x_o)$ and $I_6(x_o)$ given by Eqs. (\ref{2.26}) and (\ref{2.24}),
respectively. It is sometimes convenient to normalize the chemical 
potential, when negative, with respect to the bound-state energy $\epsilon_o$
of the associated two-fermion problem.
In the three-dimensional case, $\epsilon_o$ can be expressed in terms 
of $a_s$ whenever $a_s\geq0$. 
One finds\cite{R10}
\begin{equation}
	\epsilon_o = {1\over m a_s^2} \quad,
	\label{2.31}
\end{equation}
so that [cf. Eq. (\ref{2.29})] 
\begin{equation}
{\epsilon_o\over \epsilon_F} = {2\over (k_F a_s)^2 } = 
	{32 \over \pi^2} 
	{(x_o I_6(x_o)-I_5(x_o))^2 \over (x_o I_5(x_o) + I_6(x_o))^{2/3}}
	\quad .
	\label{2.32}
\end{equation}

\begin{figure}
\inseps{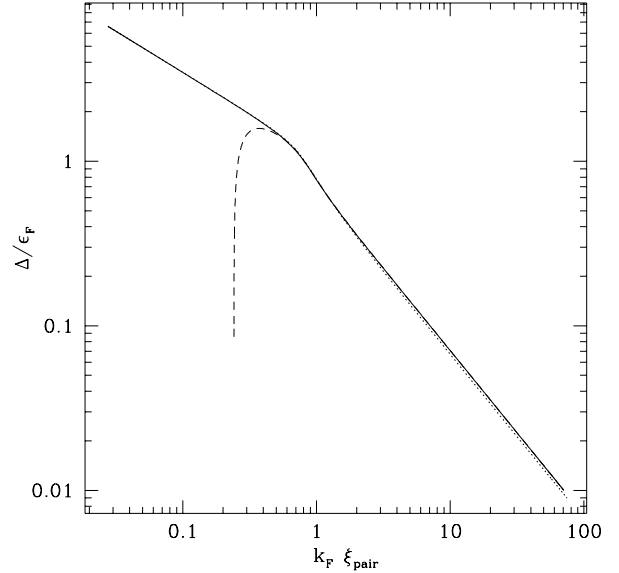}{0.4}
\caption{ $\Delta/\epsilon_F$ {\em vs} $k_F \xi_{pair}$, obtained from 
Eqs. (\ref{2.27}) and (\ref{2.30}).
Full curve: exact solution; dashed curve: BCS approximation, obtained by
including only the first two terms in Eqs. (\ref{A7}) and (\ref{A8});
dotted curve: BE approximation, obtained by including only the first 
two terms in Eqs. (\ref{A5}) and (\ref{A6}).
}
\label{f1}
\end{figure}
\begin{figure}
\inseps{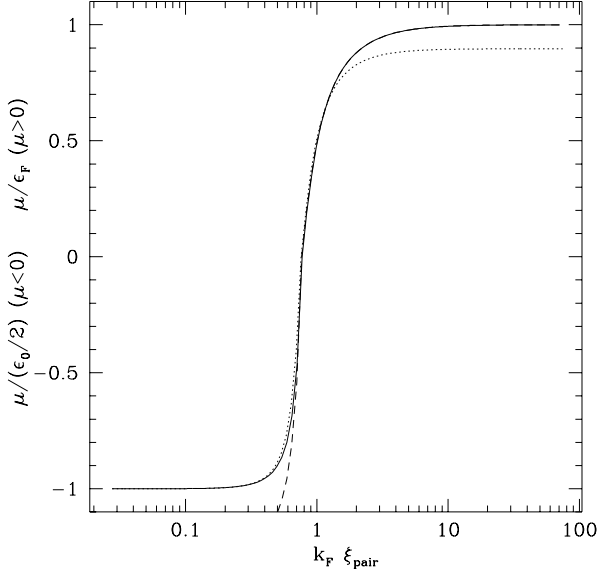}{0.4}
\caption{
$\mu/\epsilon_F$ for $\mu>0$ and $\mu/(\epsilon_o/2)$ for $\mu<0$
	{\em vs} $k_F \xi_{pair}$, obtained from Eqs. (\ref{2.28}), 
	(\ref{2.30}), and (\ref{2.32}). Conventions are as in Fig. \ref{f1}.
}
\label{f2}
\end{figure}
 
\begin{figure}
\inseps{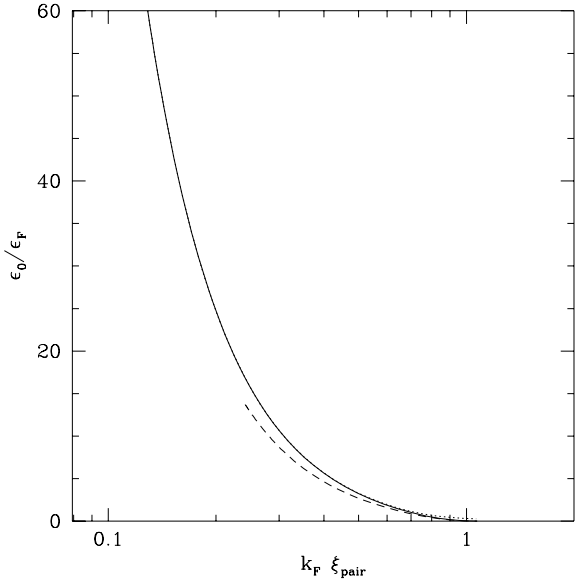}{0.4}
\caption{
$\epsilon_o/\epsilon_F$ {\em vs} $k_F \xi_{pair}$, obtained from 
	Eqs. (\ref{2.30}) and  (\ref{2.32}) when $a_s$ given by 
	Eq. (\ref{2.29}) is positive. Conventions are as in Fig. \ref{f1}.
}
\label{f3}
\end{figure}

Numerical values of the complete elliptic integrals $F$ and $E$ 
have been extensively tabulated.\cite{R12}
Otherwise, one may generate them with the required accuracy via 
Eqs. (\ref{A5})-(\ref{A8}).
In this way, the desired values of $I_5(x_o)$ and $I_6(x_o)$ 
can be obtained for given $x_o$ (or for given $k_F \xi_{pair}$ by 
inverting Eq. (\ref{2.30})). In Figs. \ref{f1}-\ref{f3} we report, 
respectively, the values of $\Delta/\epsilon_F$, $\mu/\epsilon_F$
for $\mu>0$ and $\mu/(\epsilon_o/2)$ for $\mu<0$,
and $\epsilon_o/\epsilon_F$ {\em vs\/} $k_F \xi_{pair}$ obtained by this 
procedure. 
Within numerical accuracy, all values coincide with those calculated by 
solving numerically the gap equation and the normalization condition 
(\ref{2.3}) for the limiting case of a contact potential.\cite{R10}

Besides the exact result (full curve), Figs. \ref{f1}-\ref{f3} 
show for comparison two additional curves obtained by approximating,
respectively, the elliptic integrals $F$ and $E$ by the first 
two terms of Eqs. (\ref{A5}) and (\ref{A6}) (dotted curve) 
and by the first two terms of Eqs. (\ref{A7}) and (\ref{A8}) (dashed curve).
In principle, these approximate results are expected to be 
reliable in the BE and BCS limits, in the order.
Note, however, that the BE approximate result is surprisingly accurate 
on the BCS side of the crossover.

In the BCS and BE limits the values 
of $I_5(x_o)$ and 
$I_6(x_o)$ can be obtained by retaining only a few significant 
terms in the expansions (\ref{A5})-(\ref{A8}). 
In particular, in the BCS limit $x_o\gg 1$ so that 
$\kappa^2 \simeq 1 - 1/(4 x_o^2)$ and 
\begin{equation}
	\left\{
	\begin{array}{rcl}
		I_5(x_o) &\simeq& \sqrt{x_o}\\
		I_6(x_o) &\simeq& \ln x_o/(2 \sqrt{x_o}) \quad.
	\end{array}
	\right.
	\label{2.33}
\end{equation}
Then:
\begin{equation}
\left\{
\begin{array}{rcl}
	\Delta/\epsilon_F &\simeq& 1/x_o \\
	\mu/\epsilon_F &\simeq& 1 \\
	1/(k_F a_s) &\simeq& -(2/\pi)\ln x_o \\
	k_F \xi_{pair} &\simeq& x_o/\sqrt{2} \quad . 
\end{array}
\right.
\label{2.34}
\end{equation}
In the BE limit, on the other hand, $x_o<0$ and $|x_o|\gg 1$
so that $\kappa^2 = 1/(4x_o^2)$ and 
\begin{equation}
\left\{
\begin{array}{rcl}
I_5(x_o) &\simeq& \pi/(16 |x_o|^{3/2})\\
I_6(x_o) &\simeq& \pi/(4 |x_o|^{1/2}) \quad.
\end{array}
\right.
\label{2.35}
\end{equation}
Then:
\begin{equation}
\left\{
\begin{array}{rcl}
	\Delta/\epsilon_F &\simeq& [16/(3\pi)]^{2/3}|x_o|^{1/3} \\
	\mu/\epsilon_F &\simeq& -[16/(3\pi)]^{2/3}|x_o|^{4/3} \\
	1/(k_F a_s) &\simeq& [16/(3\pi)]^{1/3}|x_o|^{2/3} \\
	\epsilon_o/\epsilon_F &\simeq& 2[16/(3\pi)]^{2/3}|x_o|^{4/3} \\
	k_F \xi_{pair} &\simeq& {1\over \sqrt{2}} 
		[16/(3\pi)]^{-1/3}|x_o|^{-2/3} \quad . 
\end{array}
\right.
\label{2.36}
\end{equation}
The limiting BCS (\ref{2.34}) and BE (\ref{2.36}) values 
coincide with those calculated previously by different methods.\cite{R10}

We mention, finally,  that another quantity which 
can be evaluated analytically at the mean-field level 
for all values of $x_o$ is the single-particle density of states.

All the above results hold for the three-dimensional system. Analogous results 
for the two-dimensional system are straightforwardly expressed in terms of 
elementary integrals and are reported for comparison in Appendix \ref{B}.

\section{Analytic results at the gaussian level}
\label{III}

Besides the quantities of Section \ref{II} defined at the mean-field 
level, additional quantities whose definition requires the introduction 
of Gaussian fluctuations can also be expressed analytically
in three dimensions at zero temperature
for the Hamiltonian (\ref{2.1}), in terms of the complete elliptic 
integrals $F$ and $E$ for {\em all\/} values of the parameter $x_o$
({\em i.e.}, following the evolution from BCS to BE). 
In particular, we shall consider the phase coherence length
$\xi_{phase}$ (associated with the spatial fluctuations of the 
superconducting order parameter) and the sound velocity $s$
(associated with the Goldstone mode of the broken symmetry).
	
	The matrix of the Gaussian fluctuations has elements\cite{R10,R15}
\begin{equation}
	{\bf \Gamma}({\bf q},\omega) = 
	\left( 
	\begin{array}{cc}
	A({\bf q}, \omega) &	B({\bf q}, \omega) \\
	B({\bf q}, \omega) &	A(-{\bf q}, -\omega) 
	\end{array}	
	\right)
	\label{3.1}
\end{equation}
 where $\omega$ is the frequency and (for a real order parameter) 
\widetext
\top{-2.5cm} 
\begin{equation}
	A({\bf q}, \omega) = {1\over \Omega} 
	\sum_{\bf k} \left({1\over 2 E_{\bf k}} - 
	{ 
	u_{\bf k}^2 u_{{\bf k}-{\bf q}}^2 
	\over 
	E_{\bf k} + E_{{\bf k}-{\bf q}} - \omega - i\eta
	}
	-
	{ 
	v_{\bf k}^2 v_{{\bf k}-{\bf q}}^2 
	\over 
	E_{\bf k} + E_{{\bf k}-{\bf q}} + \omega + i\eta
	}
	\right)
	\label{3.2}
\end{equation}
\begin{equation}
	B({\bf q}, \omega) = {1\over \Omega} 
	\sum_{\bf k} 	u_{\bf k} v_{\bf k} 
			u_{{\bf k}-{\bf q}} v_{{\bf k}-{\bf q}}
	\left( 
	{ 
	1
	\over 
	E_{\bf k} + E_{{\bf k}-{\bf q}} - \omega - i\eta
	}
	+
	{ 
	1
	\over 
	E_{\bf k} + E_{{\bf k}-{\bf q}} + \omega + i\eta
	}
	\right)
	\label{3.3}
\end{equation}
\bottom{-2.5cm} 
\narrowtext
at zero temperature. 
In these expressions, $u_{\bf k}^2 = 1-v_{\bf k}^2$ 
and $\eta$ is a positive infinitesimal. 
To extract $\xi_{phase}$ and $s$ we need consider the expansion of 
$A({\bf q}, \omega)$ and $B({\bf q}, \omega)$ for small values of 
$|\bf q|$ and $\omega$, such that
\begin{equation}
	{{\bf q}^2 \over 2m} \ll \omega^* 
	\qquad {\rm and} \qquad 
	\omega \ll \omega^*
	\label{3.4}
\end{equation}
where
\begin{equation}
	\omega^* = 
	\left\{ 
	\begin{array}{ll}
		2\Delta\quad, & \mu>0\\
		2\sqrt{\Delta^2+\mu^2} \quad, & \mu<0\quad .
	\end{array}
	\right.
	\label{3.5}
\end{equation}
The result is:
\begin{eqnarray}
	A({\bf q}, \omega) 
	&=& a_0+a_1\, \omega +a_2\, {\bf q}^2 + a_3\, \omega^2 + \dots 
	\label{3.6}
	\\
	B({\bf q}, \omega) 
	&=& b_0+b_2\, {\bf q}^2 + b_3\, \omega^2 + \dots \quad,
	\label{3.7}
\end{eqnarray}
with
\begin{mathletters}
\begin{eqnarray}
	a_0 &=& {1\over \Omega} \sum_{\bf k} 
		{\Delta^2 \over 4 E_{\bf k}^3}
	\label{3.8a}\\
	a_1 &=& -{1\over \Omega} \sum_{\bf k} {\xi_{\bf k}\over 4 E_{\bf k}^3}
	\label{3.8b}\\
	a_2 &=& {1\over \Omega} \sum_{\bf k} 
	{1\over 32m} 
	\left\{
		{\xi_{\bf k}(2 \xi_{\bf k}^2 - \Delta^2) \over E_{\bf k}^5}
		+ {{\bf k}^2 \over d m} 
		{\Delta^2(8 \xi_{\bf k}^2 + 3 \Delta^2) \over E_{\bf k}^7}
	\right\}\nonumber \\ && 
	\label{3.8c}\\
	a_3 &=& -{1\over \Omega} \sum_{\bf k} 
		{2 \xi_{\bf k}^2+\Delta^2 \over 16 E_{\bf k}^5}
	\label{3.8d}
\end{eqnarray}
\label{3.8}
\end{mathletters}
and 
\begin{mathletters}
\begin{eqnarray}
	b_0 &=& a_0
	\label{3.9a}\\
	b_2 &=& {1\over \Omega} \sum_{\bf k} 
	{1\over 32m} 
	\left\{
		-{3\xi_{\bf k} \Delta^2 \over E_{\bf k}^5}
		+ {{\bf k}^2 \over d m} 
		{\Delta^2(2 \xi_{\bf k}^2 - 3 \Delta^2) \over E_{\bf k}^7}
	\right\} \nonumber \\
	&& 
	\label{3.9b}\\
	b_3 &=& {1\over \Omega} \sum_{\bf k} 
		{\Delta^2  \over 16 E_{\bf k}^5}
	\label{3.9c}
\end{eqnarray}
\label{3.9}
\end{mathletters}
($d$ being the dimensionality). In particular, to determine $\xi_{phase}$
one has to consider the quantity\cite{R10}
\begin{equation}
	A({\bf q}, \omega=0)+B({\bf q}, \omega=0) = 2 a_0 +(a_2+b_2){\bf q}^2
	+ \dots
	\label{3.10}
\end{equation}
from which
\begin{equation}
	\xi_{phase}^2 = {a_2+b_2 \over 2 a_0} \quad .
	\label{3.11}
\end{equation}
To determine the sound velocity $s$ one has instead to consider 
the full determinant
\begin{eqnarray}
	\lefteqn{A({\bf q}, \omega)A({\bf q}, -\omega) - B({\bf q}, \omega)^2
	=}\nonumber \\
	&& 2 a_0(a_2-b_2) {\bf q}^2 +[2 a_0(a_3-b_3)-a_1^2]\omega^2
	+\dots \label{3.12}
\end{eqnarray}
which vanishes for $\omega=\omega({\bf q}) = s|{\bf q}|$, with
\begin{equation}
	s^2 = {2 a_0(a_2-b_2) \over 2 a_0(b_3-a_3)+a_1^2}
	\quad . 
	\label{3.13}
\end{equation}

	We are left with evaluating the integrals 
entering Eqs. (\ref{3.11}) and (\ref{3.13}). In {\em three\/} 
dimensions we obtain:
\widetext
\top{-2.5cm} 
\begin{eqnarray}
	a_0 &=& {m \over (2\pi)^2} (2m \Delta)^{1/2} I_5(x_o)\quad,
	\label{3.14} \\
	a_1 &=& -{m \over (2\pi)^2} 
	\left({2m \over \Delta}\right)^{1/2} I_6(x_o)
	\quad, \label{3.15}\\
	a_2+b_2 &=&
	{1\over (2\pi)^2} {1\over 4}\left(2m \over \Delta\right)^{1/2}
	\left[\int_0^\infty\!\!\!\! dx {x^2 \xi_x^3 \over E_x^5}
		-2 \int_0^\infty \!\!\!\! dx {x^2 \xi_x \over E_x^5}
		+ {10 \over 3} \int_0^\infty \!\!\!\! dx {x^4 \xi_x^2 \over E_x^7}
	\right]\quad, 
	\label{3.16} \\
	a_2-b_2 &=& 
	{1\over (2\pi)^2} {1\over 4} \left({2m \over \Delta}\right)^{1/2}
	\left[I_6(x_o) + 2 \int_0^\infty\!\!\!\! dx {x^4 \over E_x^5}\right]\quad,
	\label{3.17} \\
	b_3-a_3 &=& {1\over (2\pi)^2} {1\over 4} 
	\left({2m\over \Delta}\right)^{3/2} I_5(x_o)\quad ,
	\label{3.18}
\end{eqnarray}
\bottom{-2.5cm} 
\narrowtext
where use has been made of the notation (\ref{2.7}) and of the 
integrals (\ref{2.20}) and (\ref{2.21}). 
The four new integrals appearing in Eqs. (\ref{3.16}) and 
(\ref{3.17}) can also be expressed as linear combinations 
of $I_5(x_o)$ and $I_6(x_o)$. Integrations by parts and 
simple manipulations lead to:
\begin{eqnarray}
	\int_0^\infty\!\!\!\!dx {x^2 \xi_x \over E_x^5}
	&=& 
	{1\over 6} \int_0^\infty {1\over E_x^3} 
	= 
	{1\over 6(1+x_o^2)} 
	\left[
	\int_0^\infty\!\!\!\! dx {x^4 \over E_x^3}
	\right.\nonumber \\
	&& \qquad\left.
	-2 \int_0^\infty\!\!\!\! dx
	{x^2 \xi_x \over E_x^3} + \int_0^\infty \!\!\!\! dx {1\over E_x}
	\right]
	\nonumber \\
	&=&
	{x_o I_5(x_o) +I_6(x_o) \over 6(1+x_o^2)}
	\quad ;
	\label{3.19}
	\\
	\int_0^\infty\!\!\!\! dx {x^2 \xi_x^3 \over E_x^5}
	&=& 
	I_6(x_o) - \int_0^\infty\!\!\!\! dx {x^2 \xi_x \over E_x^5}
	\quad ;
	\label{3.20}
	\\
	\int_0^\infty \!\!\!\! dx {x^4\over E_x^5} 
	&=& 
	\int_0^\infty \!\!\!\! dx {x^2 \xi_x \over E_x^5}
	+ x_0 \int_0^\infty\!\!\!\! dx {x^2 \over E_x^5} 
	\nonumber \\
	&=& (1+x_o^2) \int_0^\infty\!\!\!\! dx {x^2 \xi_x \over E_x^5} 
	+ {x_0 \over 2} I_5(x_o)\quad ;
	\label{3.21}
	\\
	\int_0^\infty \!\!\!\! dx {x^4 \xi_x^2 \over E_x^7} 
	&=& 
	{3\over 10} \int_0^\infty\!\!\!\! dx {x^2 \xi_x\over E_x^5}
	+ {1\over 5} \int_0^\infty \!\!\!\! dx {x^4 \over E_x^5}\quad .
	\label{3.22}
\end{eqnarray}

In conclusion we obtain:
\begin{eqnarray}
\lefteqn{a_2+b_2 = {1\over (2\pi)^2} {1\over 12} 
	\left({2m \over \Delta}\right)^{1/2}} \nonumber \\
	&& \times
	\left\{2I_6(x_o)+{(1+4x_o^2) \over 3(1+x_o^2)} 
	[I_6(x_o)+x_o I_5(x_o)]\right\}
	\label{3.23}
\end{eqnarray}
and
\begin{equation}
	a_2-b_2 
	= {1\over (2\pi)^2} {1\over 3} 
	\left({2m \over \Delta}\right)^{1/2}
	\left\{I_6(x_o)+x_o I_5(x_o)\right\} \quad.
	\label{3.24}
\end{equation}
Equation (\ref{3.11}) thus reduces to
\begin{eqnarray}
	\lefteqn{
	(k_F \xi_{phase})^2 = {\epsilon_F \over \Delta} 
	{1 \over 12 I_5(x_o)}}\nonumber \\
	&& \times 
	\left\{2I_6(x_o)+{(1+4x_o^2) \over 3(1+x_o^2)} 
	[I_6(x_o)+x_o I_5(x_o)]\right\}
	\label{3.25}
\end{eqnarray}
with $\epsilon_F/\Delta$ given by Eq. (\ref{2.27}), while Eq. (\ref{3.13})
becomes
\begin{equation}
\left({s\over v_F}\right)^2 = 
	{1\over 3} {\Delta \over \epsilon_F}
	{ I_5(x_o)(I_6(x_o)+x_o I_5(x_o)) \over I_5(x_o)^2+I_6(x_o)^2}
	\label{3.26}
\end{equation}
where $v_F = k_F/m$.

\begin{figure}
\inseps{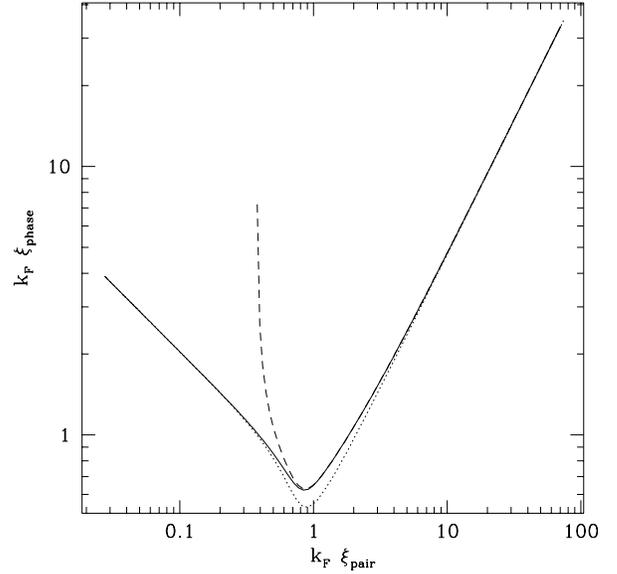}{0.4}
\caption{
 $k_F \xi_{phase}$ {\em vs} $k_F \xi_{pair}$, obtained from 
	Eqs. (\ref{2.30}) and  (\ref{3.25}). 
Conventions are as in Fig. \ref{f1}.
}
\label{f4}
\end{figure}

\begin{figure}
\inseps{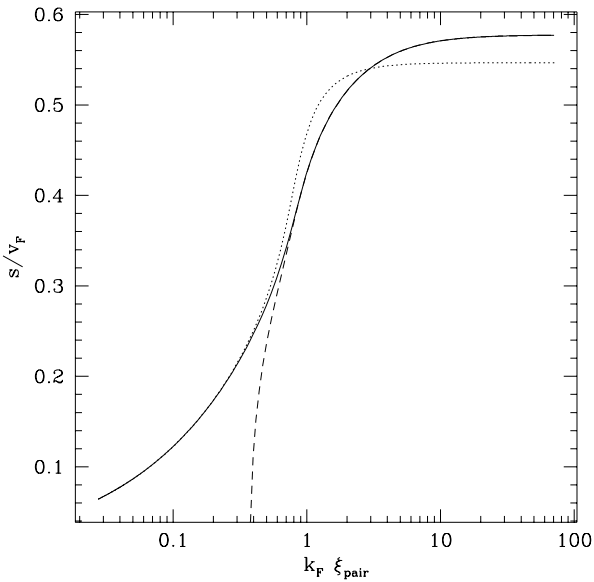}{0.4}
\caption{
 $s/v_F$ {\em vs} $k_F \xi_{pair}$, obtained from 
	Eqs. (\ref{2.30}) and  (\ref{3.26}).  
	Conventions are as in Fig. \ref{f1}.
}
\label{f5}
\end{figure}

The expressions (\ref{3.25}) and (\ref{3.26}) provide the desired 
{\em analytic\/} expressions of $k_F \xi_{phase}$ and $s/v_F$ 
for {\em all\/} values of $x_o$. Using again Eqs. (\ref{2.24}) and 
(\ref{2.26}) and 
the tabulated values of the elliptic integrals, we report in 
Figs. \ref{f4} and \ref{f5} the values  of $k_F \xi_{phase}$ 
and  $s/v_F$, respectively, {\em vs\/} $k_F \xi_{pair}$.
To within numerical accuracy, we reproduce 
in this way the numerical results of 
Ref. \onlinecite{R10} for $k_F \xi_{phase}$. 
Note, again, that the BE approximation (dotted curve) 
is surprisingly accurate even on the BCS side.

In the BCS and BE limits one can use the approximate values (\ref{2.33})
and (\ref{2.35}), in the order, for the integrals $I_5(x_o)$ and $I_6(x_o)$.
This gives:
\begin{equation}
\left\{
	\begin{array}{rcl}
	k_F \xi_{phase} &\simeq& x_o/3 \\
	s/v_F &\simeq& 1/\sqrt{3}
	\end{array}
\right.
	\label{3.27}
\end{equation}
in the BCS limit, and 
\begin{equation}
\left\{
	\begin{array}{rcl}
	k_F \xi_{phase} &\simeq& (3\pi |x_o|/16)^{1/3}\\
	s/v_F & \simeq & (12 \pi |x_o|)^{-1/3}
	\end{array}
	\label{3.28}
\right.
\end{equation}
in the BE limit. Note that in the BE limit the product
\begin{equation}
	s\,\xi_{phase} = {1\over m} \left({s \over v_F}\right)
	(k_F \xi_{phase}) \simeq {1\over 4m}
	\label{3.29}	
\end{equation}
is a constant and coincides with the Bogoliubov result 
$(2 m_B)^{-1}$ for composite bosons with mass $m_B = 2m$,
thus confirming the general results 
established in Ref. \onlinecite{R10} 
for the mapping onto 
a bosonic system in the strong-coupling limit. 

An additional quantity which can be evaluated 
analytically at the Gaussian level is the coefficient 
$\gamma$ of the quartic term in the dispersion relation
\begin{equation}
	\omega({\bf q})^2 = s^2 {\bf q}^2 + \gamma 
	\left({{\bf q}^2 \over 4m}\right)^2\quad ,
	\label{3.30}
\end{equation}
obtained by expanding the determinant (\ref{3.12}) 
to higher order. 
Its expression in rather involved and will not be reported here.
It is nonetheless interesting to note 
that in the BCS limit
\begin{equation}
	\gamma \simeq -{256 \over 135} (k_F \xi_{pair})^2
	\label{3.31}
\end{equation}
is negative (and large), so that the dispersion relation
\begin{equation}
	\omega({\bf q})^2 \simeq {v_F^2 {\bf q}^2 \over 3} 
		\left[ 1 - {1\over 3} {\bf q}^2 \xi_{pair}^2\right]
	\label{3.32}
\end{equation}
holds for $|\bf q|$ smaller then a critical value 
$q_c \propto 1/\xi_{pair}$.
This implies that in the BCS limit the wavelength 
of the collective mode associated with the symmetry breaking cannot be smaller
than the size of a Cooper pair.

The above results hold in three dimensions. 
In two dimensions both $\xi_{phase}$
and $s$ can be expressed in terms of 
elementary integrals, as shown in Appendix \ref{B}.

\section{Concluding Remarks}
\label{IV}

In this paper we have provided the analytic solution of the 
crossover problem from BCS to BE in the three-dimensional case 
for a system of fermions interacting via an attractive contact 
interaction in free space, at the mean-field level 
and with the 
inclusion of Gaussian fluctuations. 
Although the assumptions required to obtain our analytic 
solution might be oversimplified in applications to realistic 
systems, it could be interesting yet to compare our analytic
solution with numerical calculations describing more realistic 
cases. In particular, besides adopting a more 
sensible momentum-dependent form of the interaction potential,
the free one-particle dispersion relation ougth to be replaced
by the actual band structure of the medium.

In addition, a detailed description of the crossover 
problem from BCS to BE would require one to introduce already at the 
mean-field level the coupling with the charge degrees of freedom,
whose effects are expected to be especially important in the crossover
region of interest, intermediate between these two limits.\cite{R18}

In spite of these limitations, and considering the fact that 
analytic results for the crossover problem from BCS to BE were 
thus far limited to the two-dimensional case or to the two limits,
our analytic solution is useful  as it 
enables one to describe the crossover region  in a compact way 
with very limited effort.

\section*{acknoledgements}

One of us (F.P.) aknowledges receipt of financial support from Italian
INFM through the Unit\`a di Camerino and the Sezione Teorica.

\appendix

\section{Relevant properties of elliptic integrals}
\label{A}

In this Appendix we briefly review the properties of elliptic integrals 
that are relevant to our treatment.

The elliptic integrals of the first and second kinds with modulus $\kappa$
are defined by\cite{R12,R16}
\begin{eqnarray}
	F(\alpha,\kappa) &=& \int_0^\alpha \!\!\!\! d\varphi 
	{1\over \sqrt{1-\kappa^2 \sin^2\varphi}} 
	\label{A1} \\
	E(\alpha,\kappa) &=& \int_0^\alpha \!\!\!\! d\varphi 
	{\sqrt{1-\kappa^2 \sin^2\varphi}} 
	\label{A2}	
\end{eqnarray}
with $\kappa^2<1$. They satisfy the properties 
\begin{eqnarray}
	F(n \pi, \kappa) &=& 2n F({\pi\over 2}, \kappa) 
	\label{A3} \\
	E(n \pi, \kappa) &=& 2n E({\pi\over 2}, \kappa) 
	\label{A4} 
\end{eqnarray}
($n$ integer), and are said to be {\em complete\/} when 
$\alpha=\pi/2$. 
The complete elliptic integrals admit the following series 
representations:
\widetext
\top{-2.5cm} 
\def\theequation{A.\arabic{equation}}%
\begin{eqnarray}
	F({\pi\over 2}, \kappa) &=& 
	{\pi \over 2} 
	\left\{
		1+ \left({1\over 2}\right)^2 \kappa^2 + 
		\left({1\cdot3 \over 2\cdot 4}\right)^2 \kappa^4+
		\dots
		+\left({(2n-1)!! \over 2^n n!}\right)^2 \kappa^{2n}
		+\dots
	\right\}
	\label{A5}
\\
	E({\pi\over 2}, \kappa) &=& 
	{\pi \over 2} 
	\left\{
		1-\left({1\over 2}\right)^2 \kappa^2  
		-\left({1\cdot 3 \over 2\cdot 4}\right)^2 {\kappa^4\over 3}
		-\dots
		-\left({(2n-1)!! \over 2^n n!}\right)^2 {\kappa^{2n}\over 2n-1}
		-\dots
	\right\}
	\label{A6}
\\
	F({\pi\over 2}, \kappa) &=& 
	\ln{4\over \kappa'} + \left({1\over 2}\right)^2
		\left(\ln{4\over \kappa'}-{2\over 1 \cdot 2}\right)\kappa'^2
		+\left({1\cdot 3 \over 2\cdot 4}\right)^2 
		\left(\ln{4\over \kappa'} - {2\over 1 \cdot 2}
	-{2\over 3\cdot 4}\right)\kappa'^4+\dots
	\label{A7}
\\
	E({\pi\over 2}, \kappa) &=& 
	1 + {1\over2}
		\left(\ln{4\over \kappa'}-{1\over 1 \cdot 2}\right)\kappa'^2
		+{1^2\cdot 3 \over 2^2\cdot 4} 
		\left(\ln{4\over \kappa'} - {2\over 1 \cdot 2}
	-{1\over 3\cdot 4}\right)\kappa'^4+\dots
	\label{A8}		
\end{eqnarray}
\bottom{-2.5cm} 
\narrowtext
where $\kappa'=\sqrt{1-\kappa^2}$ is known as the complementary modulus.
The representations (\ref{A5}) and (\ref{A6}) are to be preferred 
when $\kappa^2 \ll 1$; when $\kappa^2\simeq 1$ and $\kappa'^2\ll 1$
the representations (\ref{A7}) and (\ref{A8}) are to be preferred instead.

Equations (\ref{2.24}) and (\ref{2.26}) of the text are obtained by
 adapting tabulated results and using the properties (\ref{A3}) 
and (\ref{A4}). We obtain:\cite{R17}
\begin{equation}
	\int_0^\infty\!\!\!\! dx {1 \over \sqrt{x^4+2 b^2 x^2+a^4}}
	=
	{1\over 2a}
	F(\pi,\kappa) = {1\over a} F({\pi\over 2}, \kappa)
	\label{A9}
\end{equation}
\begin{eqnarray}
	\lefteqn{
	\int_0^\infty\!\!\!\! dx 
	{x^2 \over \sqrt{x^4+2 b^2 x^2+a^4}}
	=
	{a E(\pi,\kappa) \over {2(a^4-b^4)} } 
	- {F(\pi,\kappa)\over 4a(a^2-b^2)}
}\nonumber \\
	&=&	
	{a \over a^4-b^4} 
	E({\pi\over 2}, \kappa)
	-{1\over 2a(a^2-b^2)}
	F({\pi\over 2},\kappa)
	\label{A10}
\end{eqnarray}
where $a^2>b^2>-\infty$, $a^2>0$, and $\kappa^2 = (a^2-b^2)/(2a^2)$.
Comparison of Eqs. (\ref{A9}) and (\ref{A10}) with Eqs. (\ref{2.24})
and (\ref{2.26}) leads us to identify
\begin{equation}
	b^2 = -x_o\quad,\qquad a^4=1+x_o^2\quad,\qquad
	\kappa^2 = {\sqrt{1+x_o^2}+x_o \over 2 \sqrt{1+x_o^2}}
	\label{A11}
\end{equation}
and results eventually into the right-hand sides of Eqs. (\ref{2.24})
and (\ref{2.26}).

\def\thesection{\Alph{section}}
 
\section{Two-dimensional case}
\label{B}
\def\theequation{B.\arabic{equation}}%

The analytic solution for the two-dimensional 
case has been already provided in Ref. \onlinecite{R11}. Its 
derivation is simpler than for the three-dimensional
counterpart obtained in this paper, since it can be expressed in terms
of  elementary integrals. For the sake of comparison, we 
report in this Appendix the two-dimensional solution on equal 
footing of the three-dimensional solution discussed in the text.

Quite generally, in two dimensions the bound-state energy $\epsilon_o$ 
exists for any value of the interaction strength $g$.
For the contact potential we are considering, however, 
the bound-state equation 
\begin{equation}
	-{1\over g} = {1\over \Omega} \sum_{\bf k} 
	{1\over {\bf k}^2/2+\epsilon_o}
	= {m\over 2 \pi} 
	\int_0^\infty dy {1\over 2y+\epsilon_o/\Delta}
	\label{B1}
\end{equation}
[with $y=k^2/(2m\Delta)$] needs to be suitably regularized  by
introducing an ultraviolet cutoff. This cutoff 
can, in turn, be removed from further consideration by combining 
Eq. (\ref{B1}) with the gap equation (\ref{2.2}), namely,
\begin{equation}
	-{1\over g} = {1\over 2 \Omega} \sum_{\bf k}
	{1\over \sqrt{\xi_{\bf k}^2+\Delta^2}}
	=
	{m\over 4\pi}
	\int_{-x_o}^\infty\!\!\!\! dz {1\over (1+z^2)^{1/2}}
	\label{B2}
\end{equation}
[with $z=y-x_o$ and $x_o$ given by Eq. (\ref{2.7})] 
which requires an analogous regularization. Performing the elementary 
integrations in Eqs. (\ref{B1}) and (\ref{B2}) one obtains
\begin{equation}
	{\epsilon_o \over \Delta} = \sqrt{1+x_o^2}-x_o\quad .
	\label{B3}
\end{equation}

The normalization condition (\ref{2.3}) gives further
\begin{equation}
n = {m \Delta \over 2\pi} \int_{-x_o}^\infty\!\!\!\! dz
	\left(1-{z\over \sqrt{1+z^2}}\right)
	=
	{m\Delta \over 2 \pi} \left(x_o + \sqrt{1+x_o^2}\right)
	\quad .
	\label{B4}
\end{equation}
 Since in the normal state 
$n= k_F^2/(2\pi) = m\, \epsilon_F/\pi$, Eq. (\ref{B4}) reads
\begin{equation}
	{\Delta \over \epsilon_F} = {2 \over x_o+\sqrt{1+x_o^2}}
	\quad .
	\label{B5}
\end{equation}
Multiplying at this point both sides of Eqs. (\ref{B3}) 
and (\ref{B5}) yields
\begin{equation}
	{\epsilon_o\over \epsilon_F} = 
	2 {\sqrt{1+x_o^2}-x_o \over \sqrt{1+x_o^2}+x_o}
	\quad .
	\label{B6}
\end{equation}

Finally, the pair-correlation length can be obtained 
from its definition [cf. Eq. (\ref{2.13}) for the three-dimensional
case]:
\begin{equation}
	\xi_{pair}^2 = {1\over m^2} 
	{ \displaystyle
	\int_0^\infty \!\!\!\! dk (k^3 \xi_k^2/E_k^6) 
	\over \displaystyle
	  \int_0^\infty \!\!\!\! dk (k/E_k^2)
	}
	=
	{2\over m \Delta} 
	{	
	\int_0^\infty \!\!\!\! dy (y \xi_y^2/E_y^6) 
	\over 
	  \int_0^\infty \!\!\!\! dy (1/E_y^2)
	}
	\label{B7}
\end{equation}
with the dimensionless quantities 
\begin{equation}
\left\{
\begin{array}{rclrcl}
	y &=& k^2/(2m \Delta)\quad,&
	x_o &=& \mu/\Delta \quad,
	\\
	\xi_y &=& y-x_o\quad, &
	E_y &=& \sqrt{\xi_y^2+1}
\end{array}
\right.	
\label{B8}
\end{equation}
[in the place of (\ref{2.7}) for the three-dimensional case].
The integrals in Eq. (\ref{B7}) are again elementary and give:
\begin{equation}
(k_F \xi_{pair})^2 = 
	{1\over 2}{\epsilon_F\over \Delta}
	\left[x_o+\left({2+x_o^2 \over 1+x_o^2}\right)
	\left({\pi\over 2} + \arctan x_o\right)^{-1}\right]\quad.
	\label{B9}
\end{equation}
It is then clear that $\Delta/\epsilon_F$, 
$\mu/\epsilon_F=x_o\Delta/\epsilon_F$, $\epsilon_o/\epsilon_F$,
and $k_F \xi_{pair}$ can be expressed as functions of 
$x_o$. 
[Note that no reference to the scattering amplitude $a_s$
has been given in two dimensions]. 

Alternatively, Eq. (\ref{B9}) with $\epsilon_F/\Delta$ given by 
Eq. (\ref{B5}) can be inverted to express $x_o$ 
(as well as all other quantities) as a function of $k_F \xi_{pair}$.

At the Gaussian level, all integrals entering the definitions 
(\ref{3.8}) and (\ref{3.9}) of the coefficients of the expansions
 (\ref{3.6}) and (\ref{3.7}) are elementary in {\em two\/} 
dimensions. 
We obtain:
\begin{eqnarray}
	a_0 &=& {m\over 8\pi} {\sqrt{1+x_o^2}+x_o \over \sqrt{1+x_o^2}}
	\label{B10}
	\\
	a_1 &=& -{m\over 8\pi} {1\over \Delta \sqrt{1+x_o^2}}
	\label{B11}
	\\
	a_2+b_2 &=& 
	{1\over 48\pi} {1\over \Delta}
	\left\{x_o + {x_o^4 + 3 x_o^2+1 \over (1+x_o^2)^{3/2}}\right\}
	\label{B12}
	\\
	a_2-b_2 &=& 
	{1\over 16 \pi} {\sqrt{1+x_o^2} + x_o \over \Delta}
	\label{B13}
	\\
	b_3-a_3 &=&
	{m\over 16 \pi} {1\over \Delta^2} 
	{\sqrt{1+x_o^2}+x_o \over \sqrt{1+x_o^2}}
	\quad .
	\label{B14}
\end{eqnarray}
These results give
\begin{equation}
	(k_F \xi_{phase})^2 = 
	{1\over 6} {\epsilon_F \over \Delta}
	{\sqrt{1+x_o^2} \over \sqrt{1+x_o^2}+x_o}
	\left\{x_o+{x_o^4+3x_o^2+1 \over (1+x_o^2)^{3/2}}\right\}
	\label{B15}
\end{equation}
and
\begin{equation}
	\left({s\over v_F}\right)^2 = {1\over 4} {\Delta \over \epsilon_F}
	\left(\sqrt{1+x_o^2} +x_o\right) \quad .
	\label{B16}
\end{equation}
Note that, owing to Eq. (\ref{B5}), $(s/v_F)^2=1/2$ is 
independent from $x_o$. On the other hand, for $k_F \xi_{phase}$
we obtain in the two limits:
\begin{equation}
	k_F\xi_{phase} \simeq 
	\left\{ 
	\begin{array}{ll}
	x_o/\sqrt{6} & \qquad{\rm BCS\ limit}\\
	1/\sqrt{8}    & \qquad{\rm BE\ limit}\quad,
	\end{array}
	\right.
	\label{B17}
\end{equation}
thus confirming that the BE limit depends markedly on  
dimensionality and shows a peculiar 
behavior in two dimensions.\cite{R10}
Note, however, that in the BE limit the product $s\, \xi_{phase}$
still coincides with the Bogoliubov result $(4m)^{-1}$.

\widetext
\end{document}